\documentclass{iau} 
\usepackage{graphicx}

\title[Star clusters and young populations in the dwarf Leo A galaxy] 
{Star clusters and young populations in \\ the dwarf irregular galaxy Leo A}

\author[R. Stonkut\.{e}, M. \v{C}eponis, A. Le\v{s}\v{c}inskait\.{e}, \& V. Vansevi\v{c}ius]   
{Rima Stonkut\.{e}$^{1,2}$, Marius \v{C}eponis$^1$, Alina Le\v{s}\v{c}inskait\.{e}$^1$, \\ \and  Vladas Vansevi\v{c}ius$^{1,2}$}

\affiliation{$^1$Center for Physical Sciences and Technology, \\ Saul\.{e}tekis av. 3, 10257 Vilnius, Lithuania \\ email: {\tt rima.stonkute@ftmc.lt} \\[\affilskip]
$^2$Astronomical Observatory of Vilnius University, \\ M. K. \v{C}iurlionis st. 29, 03100 Vilnius, Lithuania \\ email: {\tt vladas.vansevicius@ff.vu.lt}
}

\pubyear{2018}
\volume{344}  
\jname{Dwarf Galaxies: From the Deep Universe to the Present}
\editors{A.C. Editor, B.D. Editor \& C.E. Editor, eds.}
\begin{document}

\maketitle

\begin{abstract}
We have studied young stellar populations and star clusters in the dwarf irregular galaxy Leo~A using multicolor ($B$, $V$, $R$, $I$, $H\alpha$) photometry data obtained with the Subaru Suprime-Cam and two-color photometry results measured on archival HST/ACS $F475W$ \& $F814W$ frames. The analysis of the main sequence (MS) and blue supergiant (BSG -- ``blue loop'') stars enabled us to study the star formation history in the Leo~A galaxy during the last $\sim$200~Myr. Also, we have discovered 5 low-mass ($\lesssim$400~M$_\odot$) star clusters within the ACS field. This finding, taking into account a low metallicity environment and a yet-undetected molecular gas in Leo~A, constrains star formation efficiency estimates and scenarios. Inside the well known ``hole" in the H\,{\scshape i} column density map (\cite[Hunter et al. 2012]{Hunter12}) we found a shock front (prominent in $H\alpha$), implying an unseen progenitor and reminding the ``hole" problems widely discussed by \cite[Warren et al. (2011)]{Warren11}. 

\end{abstract}

\firstsection 

\section{Introduction}

Leo~A is an isolated dwarf irregular galaxy in the Local Group. It is a gas-rich (\cite[Young \& Lo 1996]{YoungLo96}; \cite[Hunter et al. 2012]{Hunter12}) dark-matter-dominated stellar system (\cite[Brown et al. 2007]{Brown2007}; \cite[Kirby et al. 2017]{Kirby2017}) of low metallicity (\cite[van Zee et al. 2006]{vanZee2006}; \cite[Kirby et al. 2017]{Kirby2017}). 

The present-day star formation activity is indicated by few H\,{\scshape ii} regions, while the existence of an old stellar population is proved by the detection of RR~Lyr stars (\cite[Dolphin et al. 2002]{Dolphin02}; \cite[Bernard et al. 2013]{Bernard13}). Detailed studies of stellar content in Leo~A were performed with the Hubble Space Telescope (HST) Wide Field and Planetary Camera\,2 (WFPC2) (\cite[Tolstoy et al. 1998]{Tolstoy98}; \cite[Schulte-Ladbeck et al. 2002]{Schulte-Ladbeck02}) and Advanced Camera for Surveys (ACS) (\cite[Cole et al. 2007]{Cole07}) by the imaging of the central part. The outer parts of the galaxy were studied with the Subaru Suprime-Cam by \cite[Vansevi\v{c}ius et al. (2004)]{Vansevicius04}, and with the HST Wide Field Camera\,3 (WFC3) by \cite[Stonkut\.{e} et al. (2018)]{Stonkute18}.

\section{Data}

\begin{figure}
\begin{center}
 \includegraphics[width=4.3in]{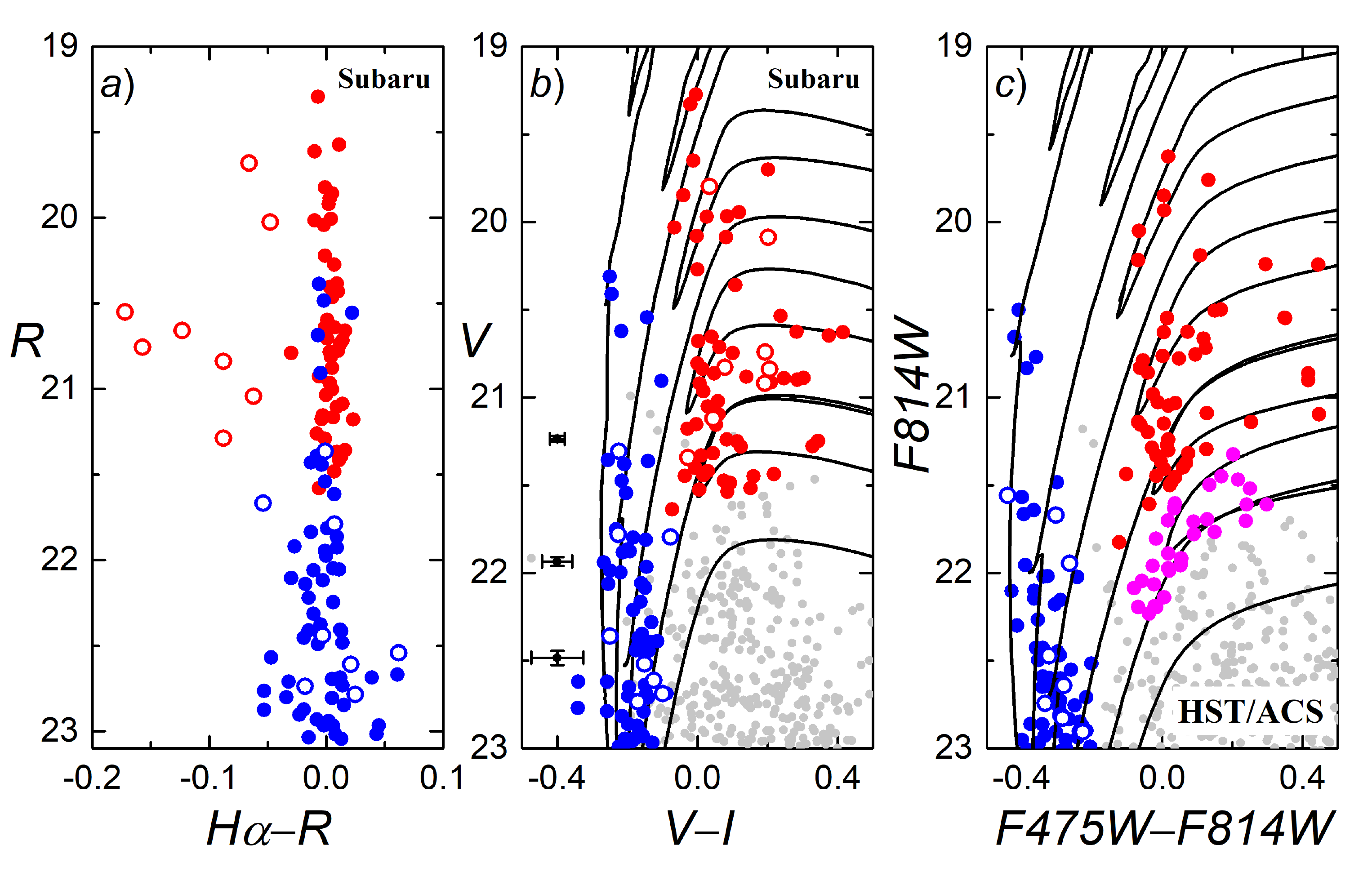} 
 \caption{CMDs of the Leo~A stars (MS -- blue; BSG -- red \& magenta). a) $R$, $H\alpha -R$ (Subaru), MS stars located in the H\,{\scshape i} ``hole" (Fig.\,\ref{fig2}a) are marked with blue open circles, BSG $H\alpha$ emission stars are marked with red open circles. b) $V$, $V-I$ (Subaru), the PARSEC isochrones (\cite[Marigo et al. 2017]{Marigo17}) of 15, 30, 55, 100, \& 160~Myr ($Z=0.0005$) are plotted. c) $F814W$, $F475W-F814W$ (HST/ACS), the isochrones of 15, 30, 55, 100, 160, \& 220~Myr ($Z=0.0005$) are plotted. All isochrones are corrected for the distance modulus of 24.5 (\cite[Dolphin et al. 2002]{Dolphin02}) and reddened assuming the MW foreground extinction of $A_{V}=0.057$, $A_{I}=0.031$, $A_{475}=0.068$, \& $A_{814}=0.032$ (\cite[Schlafly \& Finkbeiner 2011]{Schlafly2011}).}
   \label{fig1}
\end{center}
\end{figure}

We used the stellar photometry catalog (\cite[Stonkut\.{e} et al. 2014]{Stonkute14}) based on the Subaru Suprime-Cam imaging data of the Leo~A galaxy. To study young populations, we selected only bright blue stars in the color-magnitude diagrams (CMDs) shown in Fig.\,\ref{fig1}: $V<23$, $B<23$, $B-V<0.25$, $V-I<0.5$. These objects were visually inspected using multicolor Subaru and HST/ACS images. 

\begin{figure}
\begin{center}
 \includegraphics[width=4.5in]{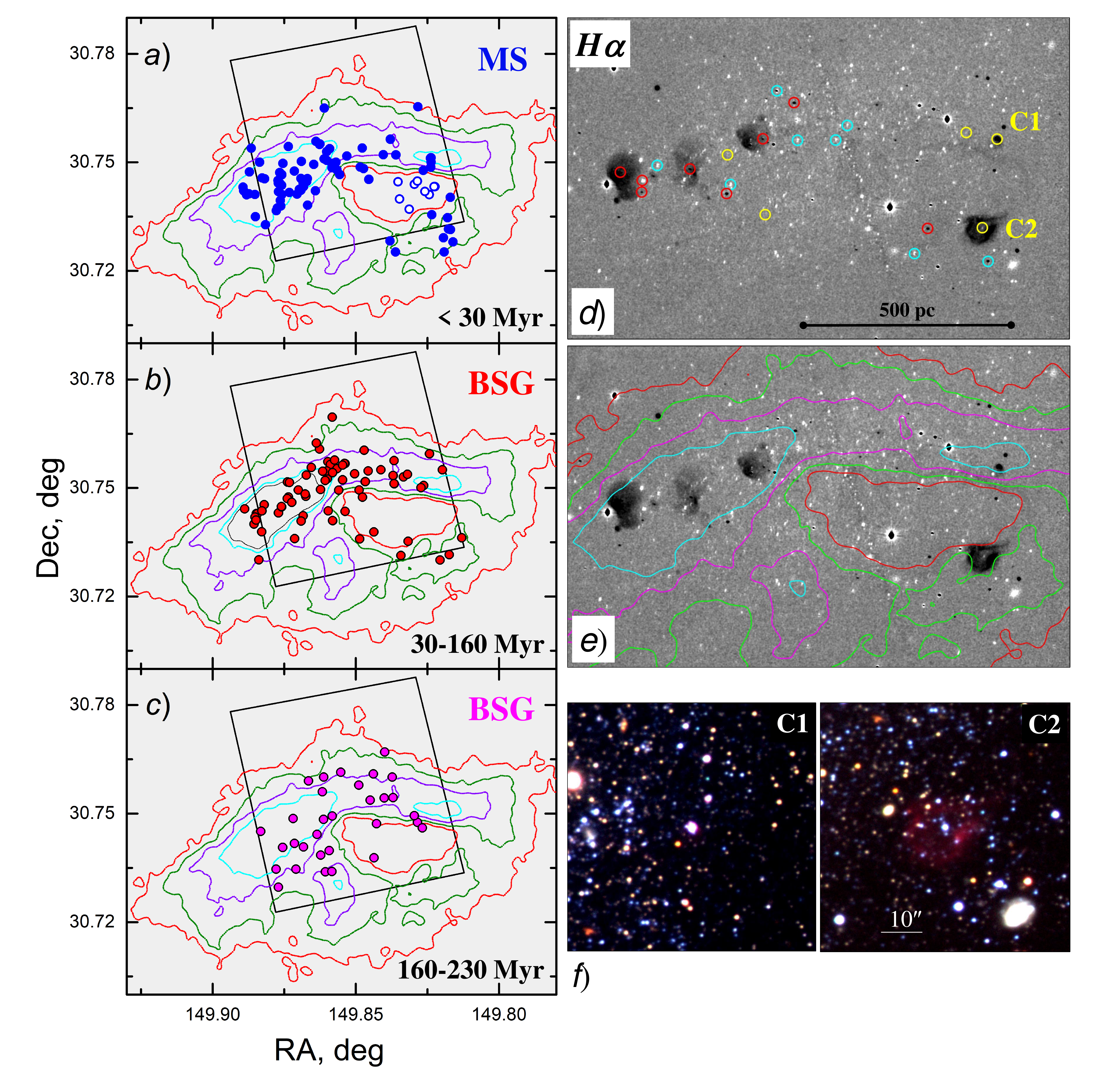} 
 \caption{a-c) MS and BSG stars -- color coding is the same as in Fig.\,\ref{fig1}. HST/ACS field is marked in black. H\,{\scshape i} column density contours ($3 \times 10^{20}$ -- red, $5 \times 10^{20}$  -- green, $7 \times 10^{20}$  -- magenta, \& $10^{21}$  -- cyan, atoms/cm$^{2}$) are shown. d \& e) the Subaru Suprime-Cam $H\alpha$ image of the Leo~A galaxy. The dark areas correspond to enhanced $H\alpha$ emission, e.g., H\,{\scshape ii} zones. d) discovered star clusters (Fig.\,\ref{fig3}, yellow open circles), BSG $H\alpha$ emission stars (Fig.\,\ref{fig1}a, red open circles); BSG stars younger than $\sim$30~Myr (cyan open circles) are shown. e) H\,{\scshape i} column density contours are plotted. f) the composite color images ($1'\times1'$) of star clusters ($H\alpha$ -- red, $V$ -- green, $B$ -- blue). North is up, East is left in all panels. 
}
   \label{fig2}
\end{center}
\end{figure}

Based on the PARSEC isochrones \cite[Marigo et al. (2017)]{Marigo17}, we estimate that selected MS stars are younger than $\sim$30~Myr and BSG star ages are from $\sim$30 to $\sim$230~Myr (Fig.\,\ref{fig1}). We selected the following objects: 88 MS stars, 9 out of them fall into the H\,{\scshape i} ``hole" area (Fig.\,\ref{fig2}a); 69 BSG stars (Fig.\,\ref{fig2}b), 8 out of them show $H\alpha$ emission (Fig.\,\ref{fig1}a); 31 BSG stars from deep HST/ACS photometry data (Figs.\,\ref{fig1}c \&\,\ref{fig2}c, filled magenta circles). 

Additionally, we used the $H\alpha$ (width of the passband $\sim$20~nm) map made by subtracting a reference frame obtained in the $R$ passband (Fig.\,\ref{fig2}d\,\&\,e). 

The integrated H\,{\scshape i} column density map (\cite[Hunter et al. 2012]{Hunter12}) was employed for the analysis of young stellar population distributions and for the ``hole's" dynamics and morphology study (Fig.\,\ref{fig2}a-c).

\begin{figure}
\begin{center}
 \includegraphics[width=4.5in]{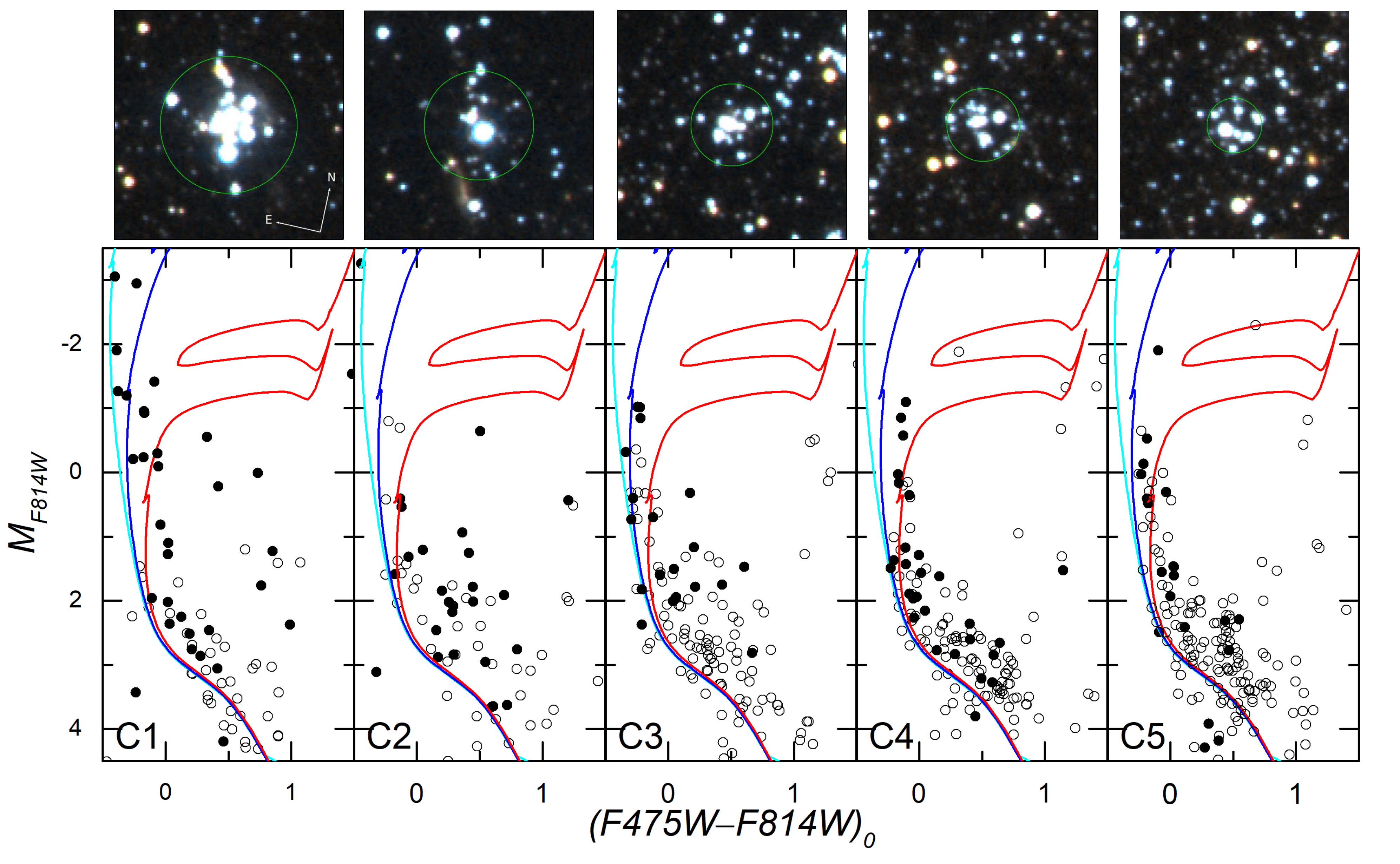} 
 \caption{The HST/ACS ($5''\times5''$) images and CMDs of the Leo~A cluster areas showing the star-like objects residing inside the green circle marking the cluster itself (filled black circles) and star-like objects residing inside the circle of a $2.5''$ radius (open circles). The PARSEC isochrones (\cite[Marigo et al. 2017]{Marigo17}) of $Z=0.0005$ metallicity and ages of 20 (cyan), 100 (blue), and 500 (red) Myr are plotted. Positions of all stars in CMDs are corrected for the distance modulus of 24.5 (\cite[Dolphin et al. 2002]{Dolphin02}) and dereddened assuming the MW foreground extinction of $A_{475}=0.068$ \& $A_{814}=0.032$ (\cite[Schlafly \& Finkbeiner 2011]{Schlafly2011}).
}
   \label{fig3}
\end{center}
\end{figure}

\section{Results and Discussion}

The analysis of the MS and BSG star distributions enabled us (by assuming there is no star migration across the disk) to visualize the 2D star formation history in the Leo~A galaxy over the last $\sim$200~Myr (Fig.\,\ref{fig2}). The analysis of young stellar populations within and around the ``hole" seen in the H\,{\scshape i} map revealed a number of interesting features: 

\begin{itemize}
\item by inspecting the H\,{\scshape i} column density distribution map, we found that the column density inside the ``hole" and in the ``hole's" walls comes as 1 to 10;
to ``inflate" the ``hole" of $\sim$500~pc in size from a single center assuming an average gas velocity dispersion of $\sim$7~km/s (\cite[Hunter et al. 2012]{Hunter12}), it would take $\sim$40~Myr; therefore, this estimate could be set as an upper age limit of the ``hole"; 
\item number densities of stars in the eastern and western parts of the ``hole" differ strongly; however, stellar populations of ages from $\sim$30 to $\sim$230~Myr are lacking in both parts (Fig.\,\ref{fig2}a-c); 
\item regions of star formation seem to avoid the ``hole" for $\sim$200~Myr; the ``hole's" western part has started to fill with a new generation of stars only recently ($<$30~Myr, Fig.\,\ref{fig2}a) and, probably, this population produces the shock front seen in the $H\alpha$ map (Fig.\,\ref{fig2}d) and induces star formation well ahead of the shock front ($\sim$20~pc); see, e.g., clusters C1 \& C2 and stars younger than $\sim$20~Myr to the South of the cluster C2;
\item remarkably, the form of the $H\alpha$ shock front closely resembles the H\,{\scshape i} column density distribution morphology (Fig.\,\ref{fig2}e), implying that the western part of the ``hole", in principle, could be swept out of H\,{\scshape i} gas by the young MS stellar population; however, the distribution of MS stars (younger than 30~Myr), projecting onto the H\,{\scshape i} ``hole's" western part, does not resemble its form and there is no increase in the number density of the young MS stars within the H\,{\scshape i} ``hole" (Fig.\,\ref{fig2}a); therefore, the problem of which objects or processes are responsible for shaping the eastern part of the ``hole" remains unsolved. 
\end{itemize}

By employing HST/ACS resolved stellar photometry data, we discovered 5 low-mass ($\lesssim$400~M$_\odot$) star clusters residing in the central part of the Leo~A galaxy (Fig.\,\ref{fig3}). All clusters are distributed around the H\,{\scshape i} ``hole" (Fig.\,\ref{fig2}d). Note, however, that these clusters are too compact to be resolved on Subaru images. Therefore, for the complete census of star clusters in the Leo~A galaxy high quality wide field HST observations are needed.  

Also, we discovered 8 BSG stars with enhanced $H\alpha$ emission (Fig.\,\ref{fig1}a \& Fig.\,\ref{fig2}d), which indicate Be stars in the Leo~A galaxy. However, 5 of them are located nearby to H\,{\scshape ii} zones and their $H\alpha$ photometry could be contaminated by a diffuse emission.
\\

$Acknowledgements$. This research was funded by a grant No. LAT-09/2016 from the Research Council of Lithuania.

\end{document}